\documentclass{aa}
\usepackage{graphicx}

\newcommand{\hcm}[1]{10$^{#1}$ cm$^{-2}$}
\newcommand{\nh}{$N_\mathrm{H}$}
\newcommand{\et}{et al.}

\begin{document}

\title{XMM-Newton EPIC Observation of SMC SNR\,0102$-$72.3\thanks{
Based on observations with XMM-Newton, an ESA Science Mission  
with instruments and contributions directly funded by ESA Member
states and the USA (NASA).}}

\author{M.\ Sasaki\inst{1} \and T.\ F.\ X.\ Stadlbauer\inst{1} \and 
F.\ Haberl\inst{1} \and M.\ D.\ Filipovi\'c\inst{1,2,3} \and
 and P.\ J.\ Bennie\inst{4}}

\authorrunning{Sasaki et al.}

%\offprints{M. Sasaki, \email{manami@mpe.mpg.de}}

\institute{Max-Planck-Institut f\"ur extraterrestrische Physik,
               Giessenbachstra{\ss}e, Postfach 1312, 85741 Garching, Germany 
           \and University of Western Sydney Nepean, P.O. Box 10,
               Kingswood, NSW 2747, Australia 
           \and Australia Telescope National Facility, CSIRO, P.O. Box
               76, Epping, NSW 2121, Australia
           \and University of Leicester, Leicester, LE1 7RH, United Kingdom}
     
\date{Received September 30, 2000; accepted November 2, 2000}

\abstract{
Results from observations of the young oxygen-rich supernova remnant 
\object{SNR\,0102$-$72.3} in the Small Magellanic Cloud during the calibration 
phase of the XMM-Newton Observatory are presented.
Both EPIC-PN and MOS observations show a ringlike structure with a radius
of $\sim$15\arcsec\ already known from Einstein, ROSAT and Chandra 
observations.  
Spectra of the entire SNR as well as parts in the eastern half were analyzed
confirming shocked hot plasma in non-uniform ionization stages as the origin 
of the X-ray emission. The spectra differ in the northeastern and the 
southeastern part of the X-ray ring, showing emission line features
of different strength. 
The temperature in the northeastern part is significantly
higher than in the southeast, reflected by the lines of higher ionization
stages and the harder continuum. Comparison to radio data
shows the forward shock of the blast wave dominating in the northern part 
of the SNR, while the southern emission is most likely produced by the 
recently formed reverse shock in the ejecta. 
In the case of the overall spectrum of \object{SNR\,0102$-$72.3}, the 
two-temperature non-equilibrium ionization model is more consistent with the 
data in comparison to the single plane-parallel shock model. The structure of 
\object{SNR\,0102$-$72.3} is complex due to variations in shock propagation 
leading to spatially differing X-ray spectra.
}

\maketitle

\section{Introduction}

\object{SNR\,0102$-$72.3} is one of the brightest X-ray sources in the Small 
Magellanic Cloud (SMC) and was discovered by the IPC instrument onboard the 
Einstein Observatory (Seward \& Mitchell \cite{SM81}). The Einstein HRI 
resolved a shell-like 
supernova remnant (SNR), and optical emission is seen from a variety of 
filaments rich in oxygen and neon (Dopita \& Tuohy \cite{DT84}) arranged in an 
incomplete clumpy ring with radius of about 12\arcsec. Enrichment in the 
element oxygen is indicative for remnants of a type II SN explosion.
The maximum velocity of the optical filaments [\ion{O}{iii}], 
not visible in H$\alpha$ (Dopita \et\ \cite{D81}), 
was found to be up to 6500\,km/s (Tuohy \& Dopita \cite{TD83}). 
The bright \ion{O}{iii} knots, assumed to be dense
ejecta clumps or Rayleigh-Taylor fingers of ejecta (density 
$\sim$1\,cm$^{-3}$), correspond in some parts well with bright X-ray
features (Gaetz \et\ \cite{G00}) but in other parts not. The estimated
ionization timescale of $\mathrm{log}(nt)<12$ 
adds to the origin of the X-ray emission
being a shock heated X-ray plasma in a state of non-equilibrium ionization
(NEI). Radio observations of \object{SNR\,0102$-$72.3} with 3\arcsec\ 
resolution in the 6\,cm  wavelength band by Amy \& Ball (\cite{AB93}) 
unveiled an outer radio shell
diameter of 40\arcsec$\pm$5\arcsec. The radio emission is found to lie
predominantly outside the bright X-ray emission but within an outer faint
X-ray rim of $\sim$44\arcsec\ (Gaetz \et\ \cite{G00}). Hayashi \et\ 
(\cite{H94}) reported on a broadband spectrum of \object{SNR\,0102$-$72.3} 
gained from the ASCA SIS instrument showing most of the emission from 
\object{SNR\,0102$-$72.3} lying in the 0.5 -- 2.3\,keV energy band. 
The most prominent line features were interpreted as He-like
K$\alpha$, H-like K$\alpha$ and  K$\beta$ emission from ionized atoms of the
elements oxygen, neon and  magnesium. Their difficulties to describe the 
overall spectrum with simple NEI
models was interpreted as an inhomogeneous abundance pattern within the
emitting plasma and possibly two components, namely forward shock in the
interstellar medium (ISM) and reverse shock in the ejecta, contributing to
the overall X-ray emission.
XMM-Newton with its high sensitivity now enables spatially resolved 
spectroscopy on various parts of \object{SNR\,0102$-$72.3}. The following 
section describes the details of the observation and the data. 
Spectra from the selected regions are presented and fitted in 
Sec.\,\ref{xrayspec}. Finally, the results are discussed in 
Sec.\,\ref{discuss}.

\section{Data}\label{data}

\begin{table*}
\caption[]{\label{obslist} XMM-Newton observations of the SMC
\object{SNR\,0102$-$72.3}} 
\begin{tabular}{ccccccc}
\hline
\noalign{\smallskip}
Orbit & Obs.\ ID & Start Time & End Time & \multicolumn{2}{c}{Pointing Direction} & Filter \\
 & & & & RA & Dec & \\
\hline\hline
\noalign{\smallskip}
65 & 0123110201 & 16.\ Apr.\ 2000, 19:06:50 & 17.\ Apr.\ 2000, 01:27:02 & 01 03 50.00 & $-$72 01 55.0 & thin \\
65 & 0123110301 & 17.\ Apr.\ 2000, 03:41:01 & 17.\ Apr.\ 2000, 09:44:33 & 01 03 50.00 & $-$72 01 55.0 & medium\\
\hline
\end{tabular}
\end{table*}

\begin{figure}
\resizebox{\hsize}{!}{\includegraphics[angle=-90]{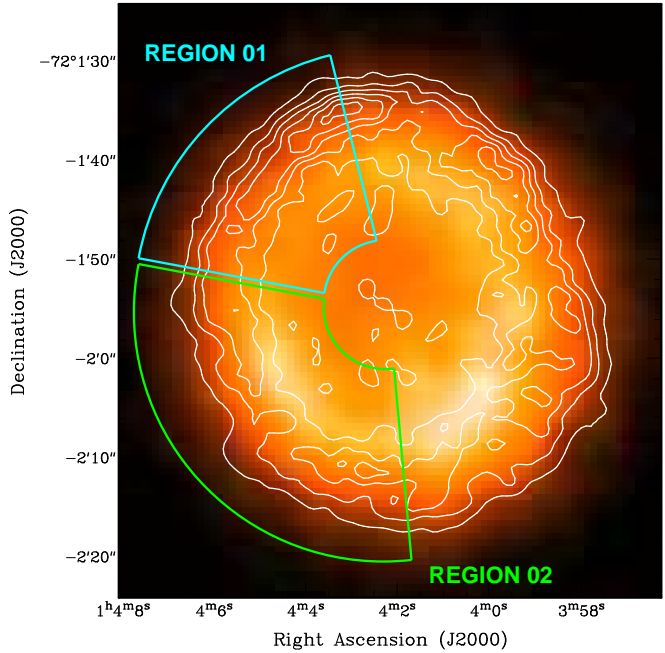}}
\caption[]{\label{mosradio} Contours of the radio image of 
\object{SNR\,0102$-$72.3} taken at ATCA superimposed on the XMM-Newton 
EPIC-MOS image (0.2 -- 3.0\,keV). 
For the radio image a combination of data presented by 
Amy \& Ball (\cite{AB93}) and data obtained from additional observations at 
ATCA is used. The radio contours are 0.1, 0.3, 0.5, 0.7, 0.9, and 
1.1\,mJy/beam. Regions selected for the spectral analysis of the EPIC-PN data
are shown.}
\end{figure}

\object{SNR\,0102$-$72.3} was observed by the XMM-Newton 
(Jansen \et\ \cite{J01}) 
in April 2000 during its calibration phase (see Table \ref{obslist}). 
There were two observations with different filters, thin and medium.
In order to study the morphologies of the SNR, XMM-Newton EPIC-MOS 
(Turner \et\ \cite{T01}) data were used. The MOS camera collected the 
emission of the complete SNR, whereas in the EPIC-PN 
(Str\"uder \et\ \cite{S01}) observations 
the SNR was split into two parts by CCD chip borders.
The EPIC-MOS data was processed through the XMM-Newton Science 
Analysis System (XMM-SAS) and events in the energy band of 0.2 -- 3.0\,keV 
were selected. 
Fig.\,\ref{mosradio} shows an overlay of the image created from the EPIC-MOS 
thin filter observation and contours of a radio image of the 4790\,MHz 
observations at the Australia Telescope Compact Array (ATCA). 

The region centered on \object{SNR\,0102$-$72.3} was observed as part of the 
ATCA mosaic observations of the SMC with a baseline of 375\,m at frequencies
of 1420 and 2370\,MHz with a corresponding angular resolution of 
$\sim$90\arcsec\ and 45\arcsec\,. ATCA observations in snap-shot mode
at 4800 and 8640\,MHz were undertaken for specific regions of interest 
including SNR 0102$-$72.3. The baseline of these observations were as well 
375\,m and
we achieved resolution of $\sim$30\arcsec\ and $\sim$15\arcsec, respectively
(Filipovi\'c \& Staveley-Smith \cite{FS98}). Amy \& Ball (\cite{AB93}) studied
the \object{SNR\,0102$-$72.3} at 4790\,Hz, but with higher resolution 
(3\arcsec). We made use of their data and merged it with our observations 
(Filipovi\'c \et\ \cite{F98} and references therein) to compensate for 
possible short space missing flux.

As can be seen in Fig.\,\ref{mosradio}, the radio emission of the SNR is 
highest in the north whereas the brightest knots in X-rays are found in the 
south. The emission from non-thermal electrons which forms the radio ring, is 
located outside the bright X-ray ring, especially in the north.
This is in good agreement with the comparison of Chandra ACIS data with the
radio image of Amy \& Ball (\cite{AB93}) by Gaetz \et\ (\cite{G00}). 
Due to higher spatial resolution
of Chandra, the ACIS image shows a detailed structure of the SNR not available
to XMM, and it was found that the radio emission is mainly located between 
the bright X-ray ring and the rim of faint X-ray emission.  

For spectral studies we analyzed the EPIC-PN data. The effective exposure 
times were 15.5\,ksec and 10.9\,ksec, respectively. We used the standard 
processed data from XMM-SAS and selected single pattern events only.
On the EPIC-PN detector the remnant was located in the CCD chips No.\ 4 and 7, 
unfortunately divided into two parts, the larger (eastern) 
part lying on chip No.\ 7. For spectral analysis of the SNR we selected a 
circular region of radius 40\arcsec\ around the center of the X-ray ring.
Events in a ring with inner radius of 3\farcm8 and outer radius of 8\farcm8 
on the chips No.\ 4 and 7, excluding out of time events, were used 
to estimate the background. In addition we 
extracted two regions of the X-ray emission with inner radius of 
8\arcsec\ and outer radius of 25\arcsec\ which is thought to arise mainly  
from the hot ejecta. The first region (hereafter REGION\,01) matches the 
northern part on the chip No.\ 7 starting from 10\degr\ to 80\degr\ 
counterclockwise from the north, and covers the faintest part of the X-ray 
ring. REGION\,02 is selected from 80\degr\ to 185\degr\ including the 
southeastern bright emission of the SNR.

\section{X-ray spectrum}\label{xrayspec}

\begin{figure}
\resizebox{\hsize}{!}{\includegraphics[angle=270,clip,bb=80 30 570 715]
{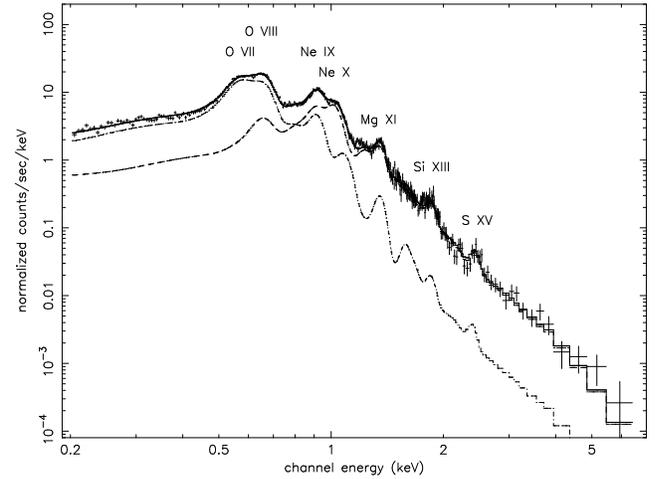}}
\caption[]{\label{specsnr} XMM-Newton EPIC-PN spectrum and fitted double
{\sf VGNEI} model of \object{SNR\,0102$-$72.3}. The dashed line shows the first
{\sf VGNEI} component (higher $\tau$) and the dash-dotted line the second 
component.}
\end{figure}

For each selected region spectra were created from both the thin and medium 
filter data separately. In order to combine both observations,
the two different spectra were fitted in XSPEC simultaneously with either 
model. In Fig.\,\ref{specsnr} the thin filter spectrum of 
\object{SNR\,0102$-$72.3} is shown.
Additionally the best double {\sf VGNEI} fit (for details see below) is 
plotted and the positions of the He-like emission lines of \ion{O}{vii}, 
\ion{Ne}{ix}, \ion{Mg}{xi}, \ion{Si}{xiii}, and \ion{S}{xv} as 
well as the H-like lines of \ion{O}{viii} and \ion{Ne}{x} are marked. 
These lines have been confirmed in the XMM-Newton RGS spectrum of the same 
pointings (Rasmussen \et\ \cite{R01}).

\begin{figure}[t]
\resizebox{\hsize}{!}{\includegraphics[angle=270,clip,bb=80 30 570 715]
{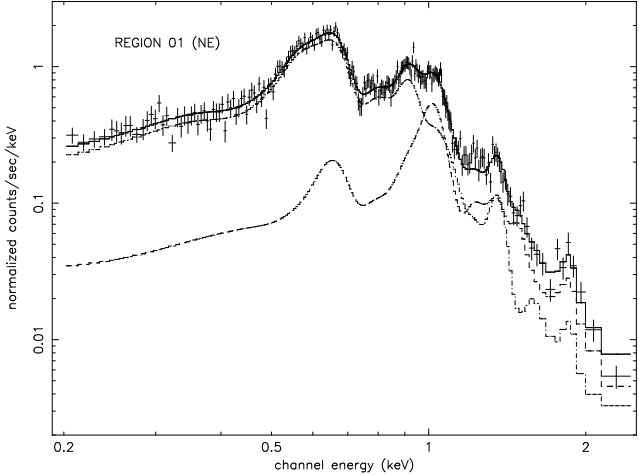}}
\caption[]{\label{specej1} Spectrum and fitted double
{\sf VGNEI} model of the northeastern part of the SNR.}
\end{figure}

\begin{figure}
\resizebox{\hsize}{!}{\includegraphics[angle=270,clip,bb=80 30 570 715]
{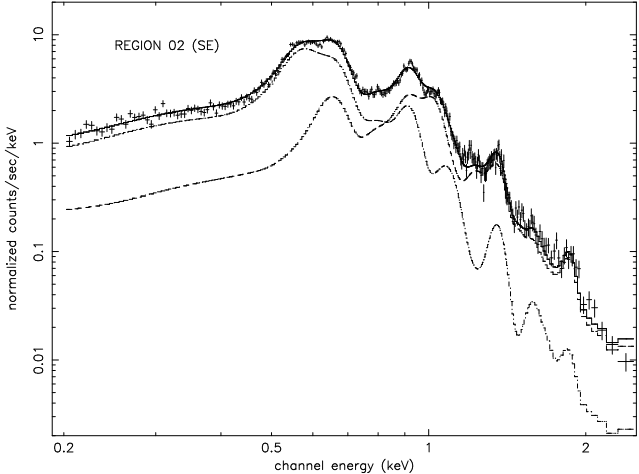}}
\caption[]{\label{specej2} Spectrum and fitted double
{\sf VGNEI} model of the southeastern part of the SNR.}
\end{figure}

Simply looking at the spectra of the northeastern and southeastern parts
(Fig.\,\ref{specej1} and Fig.\,\ref{specej2}) makes a more prominent 
line emission of the He-like ions of oxygen and neon compared to the line 
emission of the H-like ions in the southeastern spectrum obvious. 
In the spectrum of REGION\,01 the higher
ionized \ion{O}{viii} and \ion{Ne}{x} clearly are brighter. 
Even for the ions of magnesium this difference is visible. The spectrum of 
REGION\,02 shows a striking line feature of \ion{Mg}{xi} around 
1.34\,keV with a steep fall-off at higher energies. 
REGION\,01 however gives a spectrum with a less prominent 
\ion{Mg}{xi} line feature and a less steep fall-off towards the higher 
energy end, where the location of the H-like \ion{Mg}{xii} is expected 
($\sim$1.47\,keV).

Due to these He-like and H-like emission lines of oxygen and neon, as well as
other He-like lines, the spectra could not be modeled by one 
non-equilibrium ionization (NEI) model. Thus we used two generalized NEI
components with time varying temperature ({\sf VGNEI}, Borkowski \cite{B00}). 
The model {\sf VGNEI} describes a homogeneous neutral gas which was initially
cold, but is heated spontaneously. The ionized plasma is not in thermal 
equilibrium for small ionization timescales $\tau = n_\mathrm{e}t 
< 10^{12}$~s cm$^{-3}$, 
i.\ e.\ the electron temperature is lower than the ion temperature. 
The temperatures vary with time. Beside the parameters temperature $kT$ (keV), 
heavy-element abundances, and the ionization timescale $\tau$, 
the ionization timescale averaged temperature $\langle kT\rangle$ (keV)
is determined. For the two {\sf VGNEI} components, the abundances were tied
together, while the temperatures and the ionization timescales remained free 
parameters.

Furthermore we applied a plane-parallel shock model ({\sf VNPSHOCK}) which
takes different NEI states into account.
Essential parameters of the model are the mean shock temperature 
$kT_\mathrm{s}$ (keV), postshock electron temperature $kT_\mathrm{e}$ (keV)
immediately behind the shock front, and the ionization timescale $\tau$.
In collisionless shocks in SNRs electrons and ions are presumably not in 
thermal equilibrium. The electrons are heated by Coulomb collisions with ions 
which have higher temperatures in the postshock plasma. Far behind the shock 
front, electron temperature becomes equal to ion temperature for higher $\tau$.

\begin{table*}[t]
\caption[]{\label{parlist} Fit parameter} 
\begin{tabular}{lccccccl}
\hline
\noalign{\smallskip}
 & \multicolumn{2}{c}{REGION\,01} & \multicolumn{2}{c}{REGION\,02} & \multicolumn{2}{c}{entire SNR} &  \\ 
\multicolumn{1}{c}{parameter} & ~~value~~ & (90\% c.r.) & ~~value~~ & (90\% c.r.) & ~~value~~ & (90\% c.r.) & \multicolumn{1}{c}{unit} \\
\hline\hline
\noalign{\smallskip}
\multicolumn{3}{l}{{\bf generalized NEI (double {\sf VGNEI})}} & & & & & \\
\noalign{\smallskip}
SMC \nh & 8.74 & (8.01 -- 9.31) & 8.13 & (1.49 -- 24.9) & 6.93 & (5.05 -- 9.57) & \hcm{20} \\
\hline
{\scriptsize\bf 1.\ NEI component:~~~~~} & & & & & & & \\
$kT$ & 1.14 & (1.10 -- 1.26) & 1.06 & (0.46 -- 1.58) & 0.80 & (0.73 -- 0.91) & keV \\
$\tau_{1}$ & 0.81 & (0.76 -- 0.83) & 1.2 & (0.30 -- 13) & 3.9 & (3.4 -- 5.0) & 10$^{11}$ s cm$^{-3}$ \\
$\langle kT\rangle$ & 2.44 & (2.34 -- 2.57) & 0.91 & (0.56 -- 1.99) & 0.71 & (0.63 -- 0.75) & keV \\
EM$_{1}$ (frac.) & 0.58 (0.68) & & 1.7 (0.76) & & 6.2 (0.84)& & 10$^{58}$ cm$^{-3}$ \\
\hline
{\scriptsize\bf 2.\ NEI component:} & & & & & & & \\
$kT$ & 4.52 & (3.25 -- 5.47) & 1.04 & (0.30 -- 1.13) & 0.79 & (0.62 -- 0.92) & keV \\
$\tau_{2}$ & 0.11 & (0.10 -- 0.12) & 0.16 & (0.01 -- 1.0) & 0.18 & (0.07 -- 0.34) & 10$^{11}$ s cm$^{-3}$ \\
$\langle kT\rangle$ & 3.13 & (2.68 -- 3.22) & 0.74 & (0.72 -- 0.76) & 0.81 & (0.26 -- 1.88) & keV \\
EM$_{2}$ (frac.) & 0.27 (0.32) & & 0.54 (0.24) & & 1.2 (0.16) & & 10$^{58}$ cm$^{-3}$ \\
\hline
\noalign{\smallskip}
Oxygen abundance & 1.6 & (1.3 -- 1.7) & 3.9 & (3.3 -- 4.1) & 4.7 & (4.4 -- 9.3) & \\
Neon & 4.2 & (4.0 -- 4.6) & 5.9 & (5.3 -- 6.2) & 7.1 & (5.9 -- 13.5) & \\
Magnesium & 2.3 & (2.1 -- 2.5) & 3.2 & (2.9 -- 3.5) & 3.0 & (2.7 -- 3.3) & \\
Silicon & 1.0 & (0.8 -- 1.3) & 0.9 & (0.7 -- 1.2) & 0.8 & (0.6 -- 0.9) & \\
Iron & 0.6 & (0.5 -- 0.7) & 0.7 & (0.5 -- 0.8) & 0.5 & (0.4 -- 0.7) & \\
\hline
\noalign{\smallskip}
red.\ $\chi^{2}$ & 1.10 & & 1.63 & & 1.76 & & \\
\noalign{\smallskip}
\hline
\hline
\noalign{\smallskip}
\multicolumn{3}{l}{{\bf plane-parallel shock ({\sf VNPSHOCK})}} & & & & & \\
\noalign{\smallskip}
SMC \nh & 0.99 & (0.00 -- 1.29) & 1.95 & (0.99 -- 2.19) & 0.37 & (0.00 -- 0.69) & \hcm{20} \\
\hline
\noalign{\smallskip}
$kT_\mathrm{e}$ & 1.15 & (0.86 -- 1.19) & 0.08 & (0.01 -- 0.18) & 0.08 & (0.01 -- 0.20) & keV \\
$kT_\mathrm{s}$ & 5.65 & (5.64 -- 5.66) & 3.76 & (3.54 -- 3.82) & 4.20 & (4.10 -- 4.30) & keV \\
$\tau$ & 1.1 & (1.0 -- 1.2) & 1.0 & (0.9 -- 1.1) & 1.1 & (1.0 -- 1.2) & 10$^{11}$ s cm$^{-3}$ \\
\hline
\noalign{\smallskip}
Oxygen abundance & 0.8 & (0.7 -- 0.9) & 1.1 & (1.0 -- 1.2) & 1.1 & (1.0 -- 1.2) & \\
Neon & 1.4 & (1.3 -- 1.5) & 2.4 & (2.3 -- 2.5) & 2.6 & (2.5 -- 2.7) & \\
Magnesium & 0.9 & (0.8 -- 1.0) & 1.3 & (1.2 -- 1.4) & 1.5 & (1.4 -- 1.6) & \\
Silicon & 0.3 & (0.2 -- 0.5) & 0.3 & (0.2 -- 0.4) & 0.4 & (0.3 -- 0.5) & \\
Iron & 0.3 & (0.2 -- 0.4) & 0.2 & (0.2 -- 0.3) & 0.3 & (0.2 -- 0.4) & \\
\hline
\noalign{\smallskip}
red.\ $\chi^{2}$ & 1.21 & & 1.84 & & 2.52 & & \\
\noalign{\smallskip}
\hline
\end{tabular}

\vspace{1mm}
Notes: 

EM$_{1}$ and EM$_{2}$ are emission measures 
${\rm EM} = \int n_\mathrm{e} n_\mathrm{H}~\rm{d}V$ of the two 
{\sf VGNEI} components. Fractional values are given in brackets. For the 
entire SNR, EM$_{1}$ and EM$_{2}$ from the EPIC-PN spectra do not correspond 
to the total values of the SNR, because a part of the emission was not 
detected due to the CCD gap. 

Abundances are relative to solar values. 

The 90\% confidence range for temperature, ionization timescale and abundances 
are given in brackets. The confidence range for abundances are calculated by 
fixing $kT$ and $\tau$ at best fit values.

\end{table*}

The absorption consists of fixed foreground absorption with galactic 
\nh\ of $5.36\times 10^{20}\,\mathrm{cm}^{-2}$ (Dickey \& Lockman \cite{DL90}) 
and an additional absorption column density which is a free fit parameter with 
fixed abundances of 0.2 typical for the interstellar gas in the SMC 
(Russell \& Dopita \cite{RD92}). The resulting parameters of the simultaneous 
two-observation-fit for the entire SNR and the two regions are given in 
Table \ref{parlist}.

In all selected regions fitting the spectrum results in an overabundance of 
oxygen, neon, and magnesium (see Tab.\,\ref{parlist}). 
Using the double {\sf VGNEI} model, two ionization timescales were determined 
differing in one order of magnitude, $\tau \simeq 10^{11}$\,s cm$^{-3}$ and 
$\tau \simeq 10^{10}$\,s cm$^{-3}$.
The lower $\tau$ component dominates in the softer part of the spectrum below
0.8\,keV. The higher ionization states are reproduced by the higher $\tau$ 
component which is prominent in the higher energy end of the spectrum.

In REGION\,01 the determined temperatures are the highest both in the double 
{\sf VGNEI} model and the {\sf VNPSHOCK} model 
with temperatures higher than 1\,keV. This is the part with 
the lowest emission along the X-ray ring. Both the shape of the spectrum
and the temperatures indicate that this region is more highly ionized than 
in the south.

In REGION\,02 and the entire SNR spectrum the temperatures are lower, but the
abundances higher than in REGION\,01. REGION\,02 includes the brightest part 
of the X-ray ring visible in the analyzed EPIC-PN observation and the 
abundances 
of oxygen, neon, and magnesium are higher than in the northeast with 3.9, 5.9,
and 3.2 relative to solar in the double {\sf VGNEI} model, respectively. 
The comparison of the spectra of the entire SNR and REGION\,02
makes clear that the temperatures and the abundances are almost the same. 
Although there is a slight difference in the ionization timescale $\tau_{1}$ 
of the higher $\tau$ component of the double {\sf VGNEI}-model,
which is higher in the entire spectrum, the observed X-ray emission of the
SNR is dominated by the emission from REGION\,02. 
Furthermore the He-like emission lines of \ion{S}{xv} could be identified 
around 2.45\,keV in the overall spectrum, since the spectrum extends up to 
6\,keV thanks to good photon statistics.
For the whole SNR the single plane-parallel shock model yields an unsatisfying
fit. This corroborates the results of ASCA observations that the X-ray emission
of the SNR originates from at least two different thermal plasma states 
(Hayashi \et\ \cite{H94}).

\section{Discussion}\label{discuss}

\object{SNR\,0102$-$72.3} is a young SNR with an estimated age of about 
1000\,yr (Tuohy \& Dopita \cite{TD83}), which is no longer in the free 
expansion phase.
Most of the X-ray emission originates from a bright ring with radius 
$\sim$14\arcsec\ and a mean FWHM of 5\arcsec\ caused by a reverse shock 
propagating through the ejecta (Gaetz \et\ \cite{G00}; Hughes \et\ \cite{H00}).
The XMM-Newton EPIC observations confirm this overall picture. 

The spectra of the northeastern and southeastern regions of the SNR can
be fitted with a two component NEI model, each component with a 
characteristic single ionization timescale. 
The emission lines of the two highest ionization stages of oxygen and 
neon, as well as the two values for $\tau$ point out, that there 
is an ongoing shock ionization. This effect differs in the northeastern and 
the southeastern part of the SNR which can be seen in the unequal spectra of 
these regions (Fig.\,\ref{specej1} and Fig.\,\ref{specej2}). 
The plasma temperature for the northeastern 
part is up to 4 times higher than in the rest of the SNR, indicating that 
the shock velocity is higher in this region ($v_\mathrm{s} \sim 
T_\mathrm{s}^{0.5}$). Though most of the X-ray emission of the SNR arises from 
the hot ejecta, in the northeastern part the forward shock of the blast wave 
propagating into the ISM becomes important, verified in radio observations 
showing synchrotron emission right behind the blast wave which can be seen in 
Fig.\,\ref{mosradio}. The X-ray emission of the entire SNR is dominated by 
the ejecta emission, which can be seen in the significant similarity between 
the overall spectrum and the spectrum of the southeastern region. 

All the spectra extracted from EPIC-PN data were better reproduced 
by the two component NEI model than in a single plane-parallel shock model. 
The two NEI components differ not only in temperatures, but much more 
significantly in the ionization timescale $\tau$. While the lower $\tau_{2}$ 
component values themselves are similar for both selected regions as 
well as for the overall spectrum, the higher $\tau_{1}$ value is the lowest in 
the northeastern region. 
In this part of the SNR, also the fractional emission measure of the 
higher $\tau_{1}$ component is smaller than in the southeastern region or the 
whole SNR. 
Since in the southeastern (and the entire SNR) the ratio of the 
bright X-ray emission originating from the inner parts of the SNR 
(i.e.\ the bright X-ray ring) to the X-rays from the outer parts is higher
(see Fig.\,\ref{mosradio}), the higher $\tau_{1}$ component can be assigned to 
the bright X-ray ring, outlining regions with higher densities. 
The implied density distribution around the SNR ring is also supported by the 
extended overall shape along the southwest to northeast axis, which was already
reported by Gaetz \et\ (\cite{G00}) and can be verified in 
Fig.\,\ref{mosradio}. 
This is indicative of a less decelerated expansion in that direction. 
Chandra observations have shown that there is evidence for 
a spatially varying distribution of the ionization stages and the existence of 
a reverse shock in the ejecta (Gaetz \et\ \cite{G00}; 
Flanagan \et\ \cite{F01}). 
The results obtained from the spectral analysis of the XMM-Newton EPIC-PN data
show that there are differences in the plasma states between various parts
of the SNR with spatial temperature and ionization stage variations, 
contributing to its complicated structure.

\begin{acknowledgements}
The authors wish to thank Andrew Rasmussen for the valuable referee report.
The XMM-Newton project is supported by the Bundesministerium f\"ur
Bildung und Forschung / Deutsches Zentrum f\"ur Luft- und Raumfahrt
(BMBF/DLR), the Max-Planck Society and the Heidenhain-Stiftung.
\end{acknowledgements}

\end{document}